\documentstyle[epsf]{EuroPhys}
\input EuroMacr
\begin{document}
\euro{}{}{}{}
\Date{}
\shorttitle{A. MIELKE: CRITICAL TEMPERATURES OF SUPERCONDUCTIVITY}

\title{Calculating critical temperatures of superconductivity from
a renormalized Hamiltonian}
\author{Andreas Mielke\inst{1}}
\institute{
     \inst{1}Institut f\"ur Theoretische Physik,
Ruprecht--Karls--Universit\"at, Philosophenweg~19,\\
69120~Heidelberg, Germany}
\rec{}{}
\pacs{
\Pacs{74}{20.Fg}{Theories and models of superconducting state}
\Pacs{11}{10.Gh}{Renormalization}
\Pacs{71}{38.+i}{Polarons and electron-phonon interactions}
      }
\maketitle
\begin{abstract}
It is shown that one can obtain quantitatively accurate values
for the superconducting critical temperature within a Hamiltonian framework.
This is possible if one uses a renormalized Hamiltonian that contains an
attractive electron--electron interaction and renormalized single
particle energies. It can be obtained by similarity renormalization
or using flow equations for Hamiltonians. We calculate the
critical temperature as a function of the coupling 
using the standard BCS--theory. For small coupling we rederive the
McMillan formula for $T_c$. We compare
our results with Eliashberg theory and with experimental data
from various materials. The theoretical results agree with the
experimental data within 10\%.
Renormalization theory of Hamiltonians provides a promising way
to investigate electron--phonon interactions in strongly correlated 
systems.  
\end{abstract}
\section{Introduction}
For more than three decades Eliashberg theory \cite{Eliashberg}
has been the standard theory to compute quantitative properties
of superconductors like the critical temperature. For a
review we refer to \cite{Scalapino,Allen82}. The basic
advantage of Eliashberg theory compared to BCS--theory \cite{BCS}
is that it provides a framework to treat
real phonon spectra and strong electron--phonon coupling.
The critical temperature of a superconductor can be calculated with
great accuracy \cite{AD}. Eliashberg theory is based on a 
description in terms of Green's functions. It becomes exact in
the limit where the electron mass is small compared to ion masses.

In the past ten years, several new developments like the discovery
of high--$T_c$ materials \cite{Bednorz} and heavy fermions (for
a recent theoretical review see \cite{Zwicknagl}) stimulated
research activities in strongly correlated systems \cite{Fulde}. 
Theories of strongly correlated systems are usually based
on a Hamiltonian description, which is the natural basis to
investigate properties of bound states. If one wants to investigate
the effect of the electron--phonon interaction in strongly 
correlated systems, one has to set up a model Hamiltonian 
that contains phononic degrees of freedom. The main problem
is that the standard Eliashberg theory cannot be applied
to such models, because it cannot be formulated in a Hamiltonian
framework. There are also other difficulties, for instance
in heavy fermion systems, where the mass of the quasi-particles
becomes comparable to the ion masses, so that Eliashberg theory
is no longer applicable. Despite years of effort, our understanding
of electron--phonon interactions in strongly correlated
systems is rather poor. The major problem is that one needs
a theoretical method based on a Hamiltonian description that
allows an accurate treatment of the electron--phonon interaction.

BCS--theory is based on a Hamiltonian approach, but it does not
contain the electron--phonon interaction explicitely. 
Most of the properties
of superconductors can be understood using BCS--theory. But
one of the problems of BCS--theory is that the 
characteristic energy scale that determines the critical
temperature $T_c$ of the superconductor or the 
gap cannot be calculated. The famous BCS--formula
$T_c=1.13\Theta \exp(-1/N(\epsilon_F)V)$,
where $\Theta$ is the Debye temperature and $V$ is
the strength of the interaction, has to
be used to determine 
the parameter $N(\epsilon_F)V$ from $T_c$.
Usually it is argued that the problems of BCS--theory are due
to the fact that retardation effects in the interaction
are neglected \cite{Allen82}. This viewpoint is possible if one 
has a theory in terms of Green's functions in mind 
like the Eliashberg--theory. In a Hamiltonian framework
one usually works with instantaneous interactions. 
From a viewpoint
based on a Hamiltonian the problem of BCS--theory is that
the effective Hamiltonian has to be obtained from
an initial Hamiltonian containing electrons and phonons.
The effective electron--electron
interaction Bardeen {\it et al.} \cite{BCS} had in mind was the
phonon--induced interaction of Fr{\"o}hlich \cite{Froehlich}
or Bardeen and Pines \cite{BP}. These interactions are constructed 
perturbatively. But regarding the energy scales in the
problem it is clear that a perturbative approach must fail.
The initial Hamiltonian is dominated by the electronic
energies, which are typically of the order of a few eV.
Further it contains phonons with an energy scale of the order
10meV. Superconductivity arises due to a marginal relevant
operator, the phonon--induced electron--electron interaction.
The relevant energy scale in this problem is set by
the critical temperature, which is typically at least an order of
magnitude smaller than the phonon energies. Thus, if one
wants to obtain accurate values for the critical temperature,
one has to resolve an energy scale which is about five
orders of magnitude smaller than the typical energy scale
in the initial Hamiltonian. This means
that one needs a renormalization procedure which allows to
calculate an effective renormalized Hamiltonian from the
initial Hamiltonian to high accuracy.

\section{Construction of the renormalized Hamiltonian}
A general renormalization procedure for Hamiltonians 
has not been available for long time, although in his
first paper on renormalization Wilson treated a Hamiltonian
problem \cite{Wilson65}. The only accurate renormalization
procedure for Hamiltonians was numerical renormalization,
which could be applied to fermionic single impurity problems.
Three years ago, a new method for calculating effective Hamiltonians has
been proposed by Wegner \cite{Wegner94}. 
It uses continuous unitary transformations
to calculate an effective Hamiltonian from a given initial one.
This method has been called {\it flow equations for Hamiltonians}.
It has been applied successfully to single impurity problems
\cite{KM1} and to dissipative quantum systems \cite{KM}. 
Independently of Wegner, Glazek and Wilson \cite{Glazek94} 
proposed a similar method, {\it similarity renormalization}. 
They use continuous unitary transformations
to renormalize a given Hamiltonian. The aim of Wilson {\it et al.}
is to treat light--front QCD \cite{Wilson94}. 
Recently Brisudov\'a {\it et al.} calculated quarkonium spectra using
this method \cite{Brisudova}.
Both methods have a large range of possible applications
in many particle physics. The aim of this work is to show
that reliable quantitative results can be obtained with 
the framework of renormalization of Hamiltonians.

Both methods, flow equations and similarity renormalization,
have been applied to the electron--phonon problem
\cite{Lenz96,Mielke97}. Lenz and Wegner used the
flow equations to calculate the effective electron--electron
interaction. Their result differs 
significantly from the one obtained
by Fr{\"o}hlich  \cite{Froehlich}
or Bardeen and Pines \cite{BP}. The interaction
within a Cooper pair they obtained has no singularity
and is attractive in the whole parameter space.
In \cite{Mielke97} I applied the similarity renormalization
scheme to the electron--phonon problem. The effective
interaction one obtains is similar to the one
calculated by Lenz and Wegner. Furthermore, the
single particle energies in the original Hamiltonian are renormalized.
I calculated the critical temperature
for a simple Einstein model. The aim of the present work is to show that
$T_c$ can be calculated for realistic phonon spectra using this method.
The results agree very well with results from Eliashberg theory 
and with experimental data. 

Our starting point is the standard Hamiltonian for the electron--phonon
problem
\begin{equation}
  H=H_{0}+H_{I}
\end{equation}
with
\begin{equation}
  H_{0}=\sum_k\epsilon_{k}:c^\dagger_k c_k:
  +\sum_q\omega_{q} :b^\dagger_q b_q:,
\end{equation}
\begin{equation}
  H_{I}=\sum_{k,q}(g_{k,q}c^\dagger_k c_{k+q} b^\dagger_q
  +g^*_{k,q}c^\dagger_{k+q} c_k b_q).
\end{equation}
$c^\dagger_k$ and $c_k$ are the creation and annihilation
operators for electrons. I have not included spin and band
indices, but this can be done without difficulty. $b^\dagger_q$
and $b_q$ are the creation and annihilation operators for phonons,
here as well different acoustical and optical branches can
be introduced. The colons denote normal ordering.
In \cite{Mielke97} I showed how this problem can be
treated using similarity renormalization. 
The basis of this approach is a 
continuous unitary transformation applied to the Hamiltonian.
The transformation can be written in the form
$\drm H_\lambda /\drm \lambda=[\eta_\lambda,H_\lambda]$. $\lambda$
is a ultra--violett cutoff, 
and the generator $\eta_\lambda$ is chosen so
that off--diagonal matrix elements vanish in $H_\lambda$
if the corresponding energy difference is larger than $\lambda$.
Details of this method in the present context 
are explained in \cite{Mielke97}, a general
description can be found in \cite{Glazek94,Wilson94}. 
During the transformation the electron--phonon coupling is eliminated
successively and an effective electron--electron interaction
is generated. The final result is an effective
Hamiltonian that contains an electronic part of the form
\begin{equation} \label{Hr}
  H^r_{\rm el}=\sum_k\epsilon^r_{k}:c^\dagger_k c_k: 
  -\frac12\sum_{k,k^\prime,q}V_{kk^\prime q}
  :c^\dagger_{k+q} c^\dagger_{k^\prime-q} c_{k^\prime} c_{k}:,
\end{equation}
a part describing the phononic degrees of freedom and 
a weak electron--phonon coupling. The remaining electron--phonon coupling
contains a small part of the initial electron--phonon coupling
and other couplings involving two or more phonons. In
\cite{Mielke97} these couplings have been neglected since they are
of higher order in the coupling constant $g_{k,q}$. In principle
it is possible to eliminate these couplings as well. 
This yields an additional contribution to the induced
electron--electron interaction that is also of higher order.

\section{Gap and critical temperature}
In the following we will analyse the properties of the electronic
subsystem described by the renormalized Hamiltonian (\ref{Hr})
using BCS--theory. To do this we use the well known BCS gap equation 
\begin{equation} \label{gap}
  \Delta_k=\sum_q\frac{V_{k,-k,q}\Delta_{k+q}}
  {2\sqrt{\epsilon_{k+q}^{r\,2}+\Delta_{k+q}^2}}\tanh
  \left(\frac{\beta}2\sqrt{\epsilon_{k+q}^{r\,2}+\Delta_{k+q}^2}\right).
\end{equation} 
It contains the interaction $V_{k,-k,q}$ of two electrons 
forming a Cooper pair. In standard BCS--theory this interaction
is often approximated by a constant in a small energy interval
around the Fermi surface. The result of Fr{\"o}hlich \cite{Froehlich} for
this interaction is attractive in a small region around
the Fermi surface, it has a divergency due to a vanishing
energy denominator and becomes repulsive in the rest of the
parameter space. In contrary, the result by Lenz and Wegner \cite{Lenz96} is
attractive in the whole parameter space, it is of the form
$V_{k,-k,q}=|{g_{k,q}}|^2\omega_q/
((\epsilon_k-\epsilon_{k+q})^2+\omega_q^2)$.
Similarity renormalization yields a similar result,
$V_{k,-k,q}=|{g_{k,q}}|^2/(|\epsilon_k^r-\epsilon_{k+q}^r|+\omega_q)$
\cite{Mielke97}.
In the isotropic case, and introducing the 
renormalized density of states $N(\epsilon)$,
the sum in (\ref{gap}) can be replaced
by an integral. Furthermore,
the electron--phonon coupling can be described by the 
standard Eliashberg function $2\alpha^2F(\omega)$ \cite{Scalapino}.
The gap equation can then be written in the form
\begin{eqnarray} \label{gap2}
  \Delta(\epsilon)&=&\int 
  d\epsilon^\prime 
  \frac{N(\epsilon^\prime)\Delta(\epsilon^\prime)}
  {2\sqrt{\epsilon^{\prime\,2}+\Delta(\epsilon^\prime)^2}}
  \tanh\left(\frac{\beta}2\
    sqrt{\epsilon^{\prime\,2}+\Delta(\epsilon^\prime)^2}
  \right)
  \nonumber \\& &
  \frac1{N(\epsilon_F)}\int_0^\infty d\omega\frac{2\alpha^2F(\omega)}
  {|{\epsilon-\epsilon^\prime}|+\omega}.
\end{eqnarray}
The factor $1/N(\epsilon_F)$ is due to the
definition of $\alpha^2F(\omega)$, which
contains a factor $N(\epsilon_F)$. 
As a further approximation we replace the electronic
density of states $N(\epsilon)$ by a constant $N_0$. Since the
electronic energies that enter in (\ref{gap}) are renormalized,
this has to be taken into account. As in Eliashberg theory
\cite{Scalapino}, similarity renormalization yields
a renormalization of the electron energies 
$\epsilon_k^r=\epsilon_k/(1+\lambda)$ \cite{Mielke97} 
close to the Fermi surface. Therefore
the renormalized density of states $N(\epsilon_F)$ 
at the Fermi surface can be replaced by
$N(\epsilon_F)=N_0(1+\lambda)$. $\lambda$ is the usual
coupling strength in the theory of superconductivity,
defined as
$\lambda=\int_0^\infty d\omega2\alpha^2F(\omega)/\omega$.
In a first step we use (\ref{gap2}) to determine $T_c$ for
weak coupling. If $T$ is close to $T_c$, one can replace
$\sqrt{\epsilon^{\prime\,2}+\Delta(\epsilon^\prime)^2}$
by $|{\epsilon^\prime}|$. We let $\epsilon=0$ and use
that $\Delta(\epsilon^\prime)\int_0^\infty d\omega 2\alpha^2F(\omega)/
(\epsilon^\prime+\omega)$ is bounded, monotonic decreasing, and nonnegative.
The second mean value theorem then yields
\begin{equation}
  \frac{\lambda}{1+\lambda}
  \int_0^{\tilde{\omega}}d\epsilon\tanh(\epsilon/2T_c)/\epsilon=1.
\end{equation}
Keeping the leading logarithmic singularity this yields
$T_c\propto\tilde{\omega}\exp(-(1+\lambda)/\lambda)$.
$\tilde{\omega}$ can be obtained if one uses the second mean
value theorem for $\tanh(\epsilon/2T_c)$ and for $\Delta(\epsilon)$.
Then one obtains $\tilde{\omega}\propto\omega_{\rm log}$;
$\omega_{\rm log}=\exp(\langle{\ln \omega}\rangle)$, where the average
$\langle{\ln \omega}\rangle$ is taken with respect to the
weight function $\alpha^2F(\omega)/\omega$.
Taking into account that the density of states $N(\epsilon)$ is not
constant, one can introduce an additional constant in the argument
of the exponential function. In this way one obtains a $T_c$
equation of the form
\begin{equation}
  \label{McM}
  T_c=c^\prime\omega_{\rm log}\exp(-c\frac{1+\lambda}{\lambda}).
\end{equation}
It has the same form as the McMillan equation \cite{McMillan}. 
For small coupling
($\lambda<1$) $T_c/\omega_{\rm log}$ does not depend on the details
of the phonon spectrum; this has also been shown in the framework
of Eliashberg theory \cite {AD}.  

In order to obtain results for $T_c$ which can be compared
with experimental data, one has to include the Coulomb repulsion.
In principle this can be done form the very beginning. 
For our purpose it is sufficient to include the pseudo Coulomb
potential \cite{Allen82,Scalapino} $\mu^*$ by hand. 
With the assumption of a constant density of states the
gap equation (\ref{gap2}) can then be written as
\begin{eqnarray} \label{gap3}
  \Delta(\epsilon)&=&\int d\epsilon^\prime 
  \frac{\Delta(\epsilon^\prime)}
  {2\sqrt{\epsilon^{\prime\,2}+\Delta(\epsilon^\prime)^2}}
  \tanh\left(\frac{\beta}2
    \sqrt{\epsilon^{\prime\,2}+\Delta(\epsilon^\prime)^2}
  \right)
  \nonumber \\& &
  \frac{1}{1+\lambda}\left(\int_0^\infty d\omega\frac{2\alpha^2F(\omega)}
    {|{\epsilon-\epsilon^\prime}|+\omega}-\mu^*\right).
\end{eqnarray}
For small coupling one can again derive a McMillan--type formula.
For small $\lambda$ the critical temperature does not
depend on the details of $\alpha^2F(\omega)$, but
only on $\omega_{\rm log}$.
To obtain accurate results,
the gap equation (\ref{gap3}) has to be solved numerically for
a given form of the phonon spectrum. 
In Fig. 1 we show numerical results for $T_c$ in units of
$\omega_{\rm log}$ as a function of $\lambda$ for various
types of spectra and $\mu^*=0.1$. The solid curve is the results for a
spectrum of lead. Here 
$\alpha^2F(\omega)$ can be well approximated by a suitable sum 
of Lorentzians \cite{Scalapino}. This has the advantage
that the $\omega$--integral in (\ref{gap3}) can be calculated 
analytically. The remaining integral can easily be calculated numerically
to obtain $T_c$. The dotted curve is the result for an
Einstein spectrum and has already been shown already in \cite{Mielke97}.
The dashed curve shows the result for a spectrum of mercury type.
We have chosen these spectra since Allen and Dynes \cite{AD} calculated
$T_c$ for the same spectra in the framework of Eliashberg theory.
This allows a direct comparison with Eliashberg theory.
The corresponding curves lie slightly above our results, the difference
is about 5\%. We have included in our plot some experimental
data also shown in \cite{AD}. A similar calculation can be done
using the effective phonon--induced electron--electron interaction
obtained by Lenz and Wegner \cite{Lenz96}. The curves differ about
2\% from the curves shown in Fig. 1. For the Einstein model,
this has already been observed in \cite{Mielke97}.
\begin{figure}
\vbox to 10cm{\vfill\centerline{\fbox{
\leavevmode
\epsfxsize=8.2cm
\epsfysize=9.8cm
\epsfbox{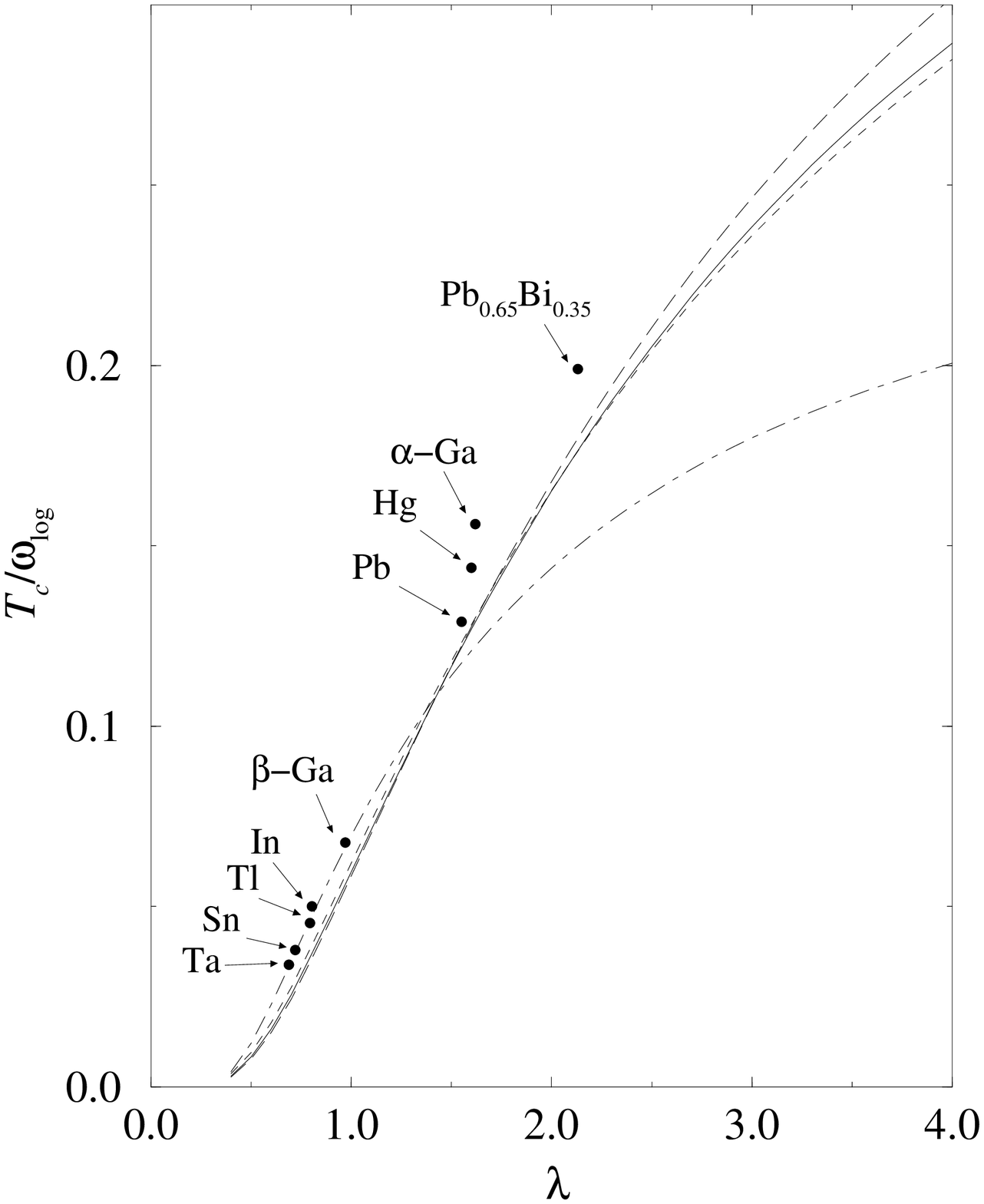}}}\vfill}
\caption{$T_c/\omega_{\rm log}$ plotted versus $\lambda$ for
$\mu^*=0.1$ and
various types of spectra. The solid line is calculated for
a lead-type spectrum, the long--dashed line for a mercury like
spectrum, and the short dashed line for an Einstein spectrum.
Also included is the McMillan curve (dot--dashed) and
some experimental data.}
\label{fig1}
\end{figure}
One notices that our results are systematically a few percent
too small compared to the results from Eliashberg theory or 
to experimental data. The reason is probably that we neglected
higher order terms in the effective Hamiltonian that describe an
interaction between electrons and two phonons. 
Treating these terms in the same way as the electron--phonon interaction,
one obtains an additional contribution to the effective electron--electron
interaction, which is of fourth order in the initial electron--phonon
coupling. This contribution is again attractive and will lead
to a somewhat higher value of the critical temperature. 

\section{Conclusions}
Our result shows that similarity renormalization or flow equations
for Hamiltonians yield an effective Hamiltonian that contains
the correct energy scale. The quantitative results are comparable to
Eliashberg theory. But these methods work in a Hamiltonian framework,
using continuous unitary transformations, and they do not rely
on special properties like a small ratio of electron mass to ion masses.
Therefore one may hope that the analytical treatment of electron--phonon
interactions in strongly correlated systems is as well possible 
using these methods. 

From the viewpoint of Eliashberg theory or of a field theoretical
approach based on a Lagrangian, where the phonons can be integrated out
explicitely, a problem may still be that the effective interaction
we calculated contains no retardation effects. But it should be
pointed out that a quantity like $T_c$ or the energy gap is not
a dynamical quantity, but a spectral property of the Hamiltonian.
Dynamical correlation functions of some observable can be calculated
as well using continuous unitary transformations. One has to take
the transformation of the observable into account. This has been
shown in a recent investigation of dissipative quantum systems
in a Hamiltonian framework \cite{KM}. Using flow equations for
Hamiltonians one can obtain accurate quantitative results for 
dynamical low temperature correlation functions as well. 

\stars
I gratefully acknowledge useful discussions with Franz Wegner,
Stefan Kehrein, Stan Glazek, and Peter Lenz.

\end{document}